\documentclass[letterpaper,10pt]{article} 

\usepackage{osameet3} 

\usepackage{amsmath,amssymb}
\usepackage[colorlinks=true,bookmarks=false,citecolor=blue,urlcolor=blue]{hyperref} 
\usepackage{amsmath,amssymb} 
\usepackage[bookmarks=false]{hyperref} 
\usepackage{layouts}
\usepackage{subcaption}
\usepackage{pgfplots, setspace}
\begin{document}


\title{Model-Based Deep Learning of Joint Probabilistic and Geometric Shaping for Optical Communication}

\author{Vladislav Neskorniuk\textsuperscript{1,3}, Andrea Carnio\textsuperscript{2}, Domenico Marsella\textsuperscript{2},\\Sergei K. Turitsyn\textsuperscript{3}, Jaroslaw E. Prilepsky\textsuperscript{3},  Vahid Aref\textsuperscript{1}}
\address{\textsuperscript{1}Nokia, 70469 Stuttgart, Germany\quad\textsuperscript{2}Nokia, Vimercate 20871, Italy\\
\textsuperscript{3}Aston Institute of Photonic Technologies, Aston University, B4 7ET Birmingham, UK}
\email{v.neskorniuk@aston.ac.uk }
\copyrightyear{2022}

\begin{abstract}
    Autoencoder-based deep learning is applied to jointly optimize geometric and probabilistic constellation shaping for optical coherent communication. The optimized constellation shaping outperforms the 256 QAM Maxwell-Boltzmann probabilistic distribution with extra 0.05~bits/4D-symbol mutual information for 64 GBd transmission over 170 km SMF link.
\end{abstract}

\section{Introduction}
Geometric (GS) and probabilistic (PS) constellation shaping, i.e. the optimization of, respectively, locations and occurrence probabilities of the constellation points, can improve considerably the transmission rate of coherent fiber-optic communication systems. While for linear channels, the family of Maxwell-Boltzmann (MB) distributions leads to the optimal PS~\cite{kschischang1993optimal}, finding the optimal shaping for nonlinear-dispersive optical fiber channels is a subtle problem~\cite{fehenberger2016probabilistic}. Usually, a time-consuming numerical statistical modeling is used for the optimization of either GS~\cite{chen2018increasing}, PS~\cite{fehenberger2016probabilistic}, or constrained joint geometric and probabilistic shaping (JS)~\cite{cai2020performance}.


End-to-end (E2E) learning~\cite{o2017introduction} offers a different solution to this problem with a gradient-based optimization of constellation shaping which does not require any  constraints on the spatial or probabilistic distributions of constellation points. Nonetheless, the previous works on E2E learning of coherent systems were focused on the optimization of GS~\cite{jones2019end, uhlemann2020deep, song2021end}. Here, we apply the E2E learning technique proposed in \cite{stark2019joint} to jointly optimize geometric and probabilistic shaping in simulation for high symbol-rate coherent transmission over single-span fiber link.
Using training through the nonlinear interference noise (NLIN) model~\cite{dar2014accumulation} in the numerical simulation, we demonstrate  that 
the optimized JS constellation leads to a mutual information (MI) gain over the MB PS of a quadarature-amplitude modulation (QAM) constellation.

\begin{figure}[b]
    \centering
    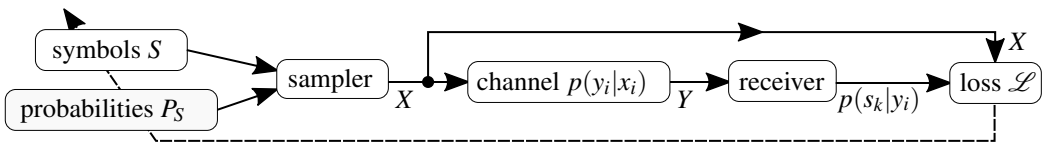
    \caption{\small The principal scheme of an end-to-end learning algorithm.}
    \label{fig:e2e-detailed}
\end{figure}

\section{Joint Learning of Geometric and Probabilistic Shaping}

In an autoencoder-based E2E learning, the trainable transmitter (TX), receiver (RX) and the channel are implemented by cascades of neural networks (NN) which are jointly trained to best reproduce the TX inputs from the RX outputs~(Fig.~1)~\cite{o2017introduction}. Following \cite{stark2019joint},
the considered E2E architecture starts with the trainable TX sampling a symbol sequence $X$. We implemented the TX utilizing a trainable sampler based on the Gumbel-Softmax trick~\cite{ stark2019joint}. The TX has two sets of trainable parameters: the set of the constellation symbols location, denoted by $S=\{s_1,s_2,...,s_N\}$, and their occurrence probabilities, denoted by $P_S=\{p_1,p_2,...,p_N\}$. Each symbol location $s_k$ and probability $p_k$ is trained separately with the only constraint on the average power: $\sum_k p_k |s_k|^2 = 1$. 

Next, a pre-defined stochastic channel model, $p(y_i|x_i)$, maps the transmitted symbols ${X = \{x_i\},\,x_i\in S}$ to the received ones, $Y=\{y_i\}$. We modeled the channel by a nonlinear interference noise (NLIN) model~\cite{dar2014accumulation}. In details, we approximated the symbol distortion as a Gaussian noise with the variance $\sigma^2 = \sigma^2_{ASE} + P^3 \left( \chi_0 + \chi_1(\mu_4-2) + \chi_2(\mu_6 - 9\mu_4 + 12) + \chi_3(\mu_4 - 2)^2 \right)$, where $\sigma^2_{ASE}$ is the variance of amplified spontaneous emission (ASE) noise injected by optical amplifiers (OAs), $P$ is the average launch power, $\mu_4,\,\mu_6$ are the 4th-order and 6th-order standardized moments of the input sequence $X$, and $\chi_0,\, \chi_1,\,\chi_2,\,\chi_3$, are the modulation-format-independent NLIN model coefficients. Then, RX maps each received symbol $y_i$ to the vector of posterior probabilities $p(s_k|y_i)$ for all $s_k\in S$, estimating how likely $s_k$ was sent when $y_i$ is received. Our RX estimates the posteriors distribution in closed form via Bayesian rule from the conditional density defining the channel $p(y_i|x_i)$ and the symbol probabilities $P_S$. As shown in \cite{stark2019joint}, the E2E MI, $I(X;Y)$, can be characterized in terms of $p(s_k|y_i)$ and $P_S$. To numerically maximize MI, the batch gradient descent algorithm with Adam optimizer is used to optimize the training parameters, i.e. $S$ and $P_S$. The gradients are computed using back-propagation through the autoenconder.

To assess the learned constellations, we generated two independent random sequences of constellation symbols sampled according to the optimized $P_S$. These sequences are the transmission data over two polarizations. The transmission is simulated using a precise Manakov equations split-step (MSS) solver.
Finally, the E2E mutual information of received symbols are calculated using Gaussian kernel density estimator which numerically estimates the MI with high accuracy.


\begin{figure}[t]
    \centering
    \includegraphics[width=\textwidth]{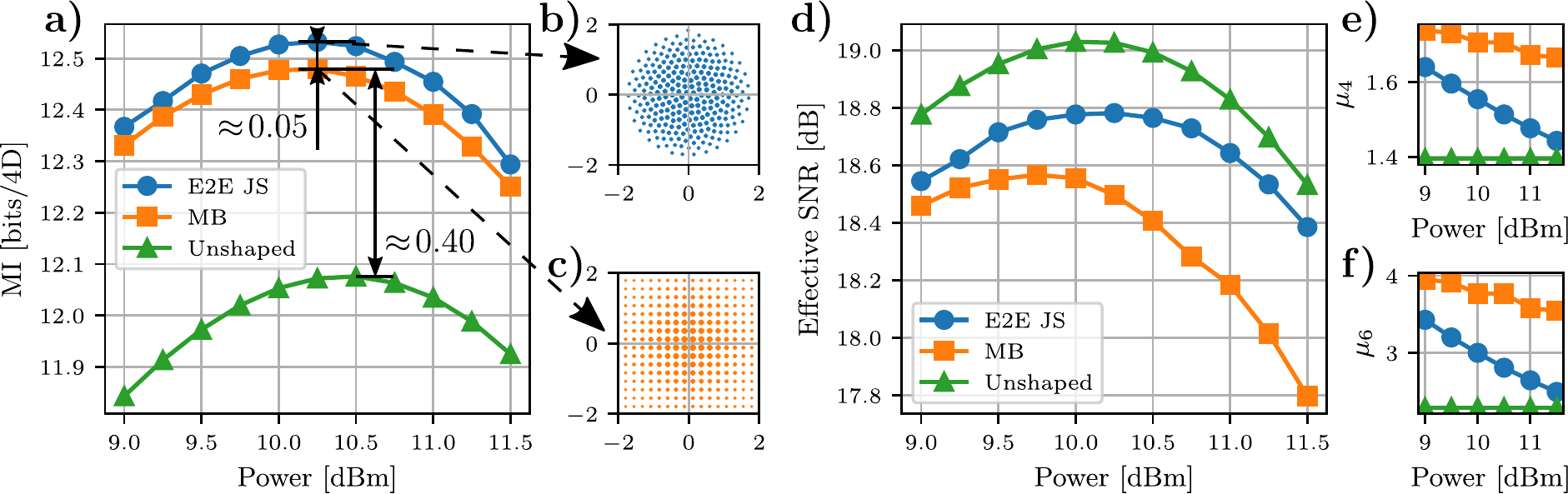}
     \caption{\small Numerical estimations of (a) mutual information (MI) over 4D symbols, (d) effective signal-to-noise ratios (SNRs), (e) $\mu_4$ 4th-order, and (f) $\mu_6$ 6th-order standardized moments for unshaped 256QAM, MB-shaped 256QAM, and E2E-learnt constellations over single-channel 64 GBd 170 km SMF link; (b) E2E-learnt and (c) MB constellations at optimal power level, marker size proportional to symbol probability $p(s_k)$. SMF parameters are taken as $D=16.8$ ps/(nm*km), $\gamma=$ 1.14 1/(W*km), $\alpha=$ 0.21 dB/km, OA noise figure = 4.5 dB. The pulses are shaped by a root-raised-cosine filter with roll-off 0.1.\vspace{-5mm}}
    \label{fig:e2e-detailed}
\end{figure}
\vspace{-1mm}
\section{Results and Conclusions}

In order to obtain a strong dependence of the nonlinear distortion on the modulation format a single-span link was studied~\cite{fehenberger2016probabilistic}. Particularly, we considered a dual-polarized single-channel 64 GBd transmission over a 170 km standard single-mode fiber (SSMF) span, followed by an ideal OA.

The effectiveness of the E2E-learned 256-symbol JS constellation compared with MB-shaped 256QAM and unshaped 256QAM was evaluated in terms of MI and reported in Figure 2. The optimization process of E2E-learned and MB constellations was done separately for each power level by considering the NLIN channel model, and the MSS solver, respectively.
The E2E-learned constellation provided a peak-to-peak MI gain of 0.05 bits/4D over MB-shaped 256QAM and 0.45 bits/4D over unshaped 256QAM. The learned JS offers an effective trade-off between linear regime performance and tolerance to nonlinear distortion: in linear regime, similarly to the MB PS, it leads to a better MI performance than the unshaped constellation, while in the nonlinear regime it produces higher effective SNR (Fig.~2d) than the MB PS, due to lower moments $\mu_4, \mu_6$ (Figs. 2e, 2f). Note that a small mismatch between the optimal launch powers in terms of SNR and MI curves is because the MI is computed by density estimator in complex domain but SNR computation neglects correlations in complex domain.


\begin{singlespace}
\footnotesize
\textit{Acknowledgements}:
This project has received funding from EU Horizon 2020 program under the Marie Sk\l{}odowska-Curie grant agreement No. 766115 (FONTE). SKT acknowledges the support of EPSRC project TRANSNET.
\end{singlespace}
\vspace{-2mm}
\normalsize

\end{document}